\documentstyle[11pt,newpasp,twoside,epsf]{article}
\newcommand{\sech}{\mbox{\rm sech}}

\begin{document}

\title{Distribution of dust and stars in the Galaxy}
\author{Ronald Drimmel}
\affil{Osservatorio Astronomico di Torino, Pino Torinese, TO 10025, Italy}

\author{David Spergel}
\affil{Princeton Observatory, Princeton, NJ}

\begin{abstract}
Using far-infrared $240 \micron$ and near-infrared K band data
from the COBE/DIRBE instrument,
we model the Galactic stellar and dust distribution.
Making the assumption that the Galaxy is transparent in the
$240 \micron$ band, the dust emission is modeled using the following
components: a warped exponential disk of scale length $0.26 R_\odot$,
a spiral arm component as mapped by HII regions,
and a feature coinciding with the local (Orion) arm.
The dust distribution is used to calculate absorption in the K band,
and the stellar emission is likewise modeled with a warped exponential 
disk, with a scale length of $0.29 R_\odot$, and a spiral arm component.
Models of the K band emission in the Galactic plane 
indicate that in this waveband a two arm spiral dominates the 
nonaxisymmetric emission.
The warp is evident in both the dust and stellar component, and
is found to start within the Solar Circle.
\end{abstract}

\section{Introduction and Data}

The emission from our own Galaxy is naturally far greater than that 
from all other galaxies combined, and is received from all directions due to
our location within it. Yet it is just this last fact which presents 
special challenges in any effort to infer the structure of our Galaxy on
a large scale, particularly of the stellar component in which most of the 
disk mass resides. From radio observations large-scale 
nonaxisymmetric structure was first seen, but only with the advent of 
infrared observations has such structure likewise been evidenced in the 
stars.

Here we model the $240\micron$ and K band emission as
detected by the COBE satellite, concentrating on the Galactic plane (GP)
emission for galactic radii $r > .5 R_\odot$, and its 
nonaxisymmetric nature. 
This contribution can be regarded as a preview of a more detailed 
exposition and analysis, incorporating the J band, that will be published 
elsewhere. Here we will principally focus on the spiral structure 
in the stellar distribution inferred from the K band alone.

\begin{figure}[ht]
\plotfiddle{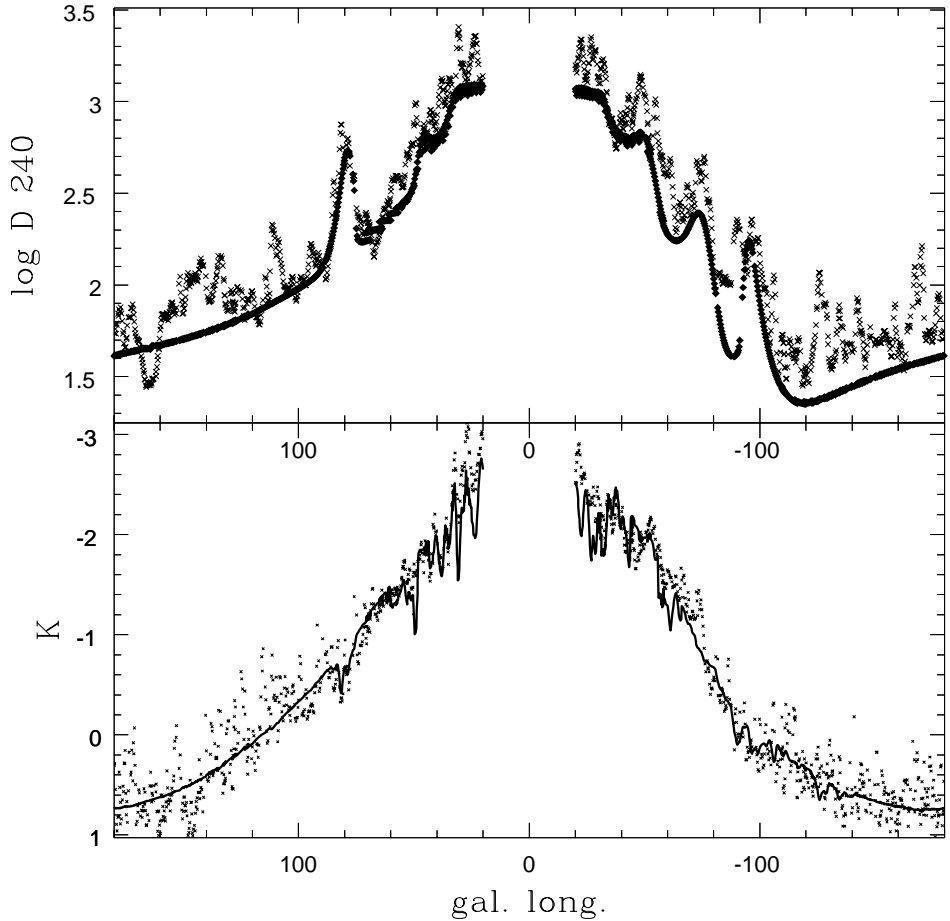}{12.0cm}{0}{65}{65}{-200}{-100}
\caption{
$240 \micron$ and K band, modeled and observed, emission profiles 
in the Galactic plane, on a logarithmic scale ($\log D$).
Data is represented by the X's.
}
\end{figure}

Galactic emission in the FIR $240\micron$ and NIR K band was
detected by the DIRBE instrument aboard the COBE satellite, from 
which 'Zodi-Subtracted Mission Average (ZSMA)' skymaps where produced
(Kelsall et al. 1998). From these maps
we take emission from galactic latitudes $|b| < 30 \deg$, excluding 
the Galactic Center region, and smaller regions centered on
the Orion Nebula, M31, M33, and the Magellanic Clouds. Excluding the 
Galactic center obviates the need to model the Galactic bulge.
In the NIR dataset point sources are also excluded, while in the FIR
the data was smoothed using a second-order two-dimensional polynomial 
applied to a 7 $\times$ 7 pixel window. This smoothing procedure introduces
no systematic redistribution of flux associated with the Galactic
plane, and was employed only to smooth over artificial negative intensities 
introduced in the data set by the subtraction of Zodiacal light.
The resulting data set consists of 173569 pixels at $240\micron$ 
and 152371 pixels in the K band.

\section{Dust distribution model}

Three density components are
used to model the dust distribution: an axisymmetric disk, a spiral
arm component, and a local feature described as a spiral segment corresponding
to the Orion arm. The axisymmetric disk component is assumed to have
an exponential radial density profile with a prescribed central hole 
and a $\sech^2$ vertical profile with a linearly flaring scale height.
The disk emissivity is derived by adopting a linear temperature gradient,
consistent with that seen in the dust associated with HI 
(Sodroski et al. 1994), with the assumption that the dust emissivity is 
$\propto \nu^2 B_\nu(T)$, $B_\nu(T)$ being Planck's function 
(Draine \& Lee 1984; Dwek 1995; Schlegel, Finkbeiner \& Davis, 1998).

The spiral arms have a double gaussian density profile in their cross section, 
an assumed geometry that is adopted from a map of HII regions 
(Georgelin \& Georgelin 1976; Taylor \& Cordes 1993), and
a quadratically flaring scale height. We also found it necessary to
apply a reduction factor to the size (width and height) and density
of the Sagittarius-Carina (Sag-Car) arm. The spiral arms are assigned a
constant emissivity (single mean temperature). The local feature
has a cylindrical gaussian cross section in density and
two associated emissivities, one for positive galactic longitudes,
the other for negative longitudes. The Sun is found in a gap 
of this local feature.
Finally, to all the density components is added 
a global warp described by
\begin{equation}
Z_{\rm w}(r,\phi) = a_{\rm w}(r-r_{\rm w})^2 \sin (\phi - \phi_{\rm w})
\end{equation}
for $r > r_{\rm w}$, where the parameters 
$a_{\rm w}$, $r_{\rm w}$, and $\phi_{\rm w}$ are adjusted.

The FIR emission is modeled on the assumption that the Galaxy is optically
thin at $240\micron$, leading to 
\begin{equation}
\label{idust}
D_{240}^{\rm mod}(l,b) = \int_{\rm los} \sum_i k_i \rho_i ds + Q_{240} 
= \sum_i D_i + Q_{240} ~,
\end{equation}
where $k_i$ and $\rho_i$ are the emissivities and densities of the $i$ 
components, and $Q_{240}$ is an isotropic term which can be 
equated with the Cosmic Infrared Background (CIB).
However, the decomposition of the flux
density into densities and emissivities cannot be done based on the 
FIR emission alone; for the adjustment to the FIR data emissivity values
are fixed. The size of the emitting regions are also indeterminate
and the model is assigned a scale by defining length units such
that $R_\odot = 1$. 

The parameters of the dust model were adjusted using a $\chi^2$ minimization
routine against the DIRBE $240 \micron$ data. The resulting fit in the 
GP is shown in Figure 1 while
emission profiles at other galactic latitudes is shown in Figure 2.
(The apparent increase in
the residuals with latitude are an effect of the logarythmic scale.)
As can be seen in these emission profiles, while the model is unable to
reproduce small scale features, the major nonaxisymmetric features
are accurately placed. While the fit in the plane itself is not exceptional,
it is a significant improvement over the axisymmetric models used in 
the past. In fact, the importance of the nonaxisymmetric features in the 
dust distribution is emphasized as one moves out of the GP,
where these features clearly dominate the emission profiles. Here the 
ability of the model to reproduce the emission profiles is much more 
impressive. The scale length resulting from the adjustment is $0.26 R_\odot$,
while the scale height is found to be 147\,pc (assuming $R_\odot = 8$\,kpc).
We also mention that a CIB of 1.07\,MJy/sr is recovered, 
consistent with other estimates (Hauser et al. 1998, Lagache et al. 1999).

\begin{figure}[ht]
\plotone{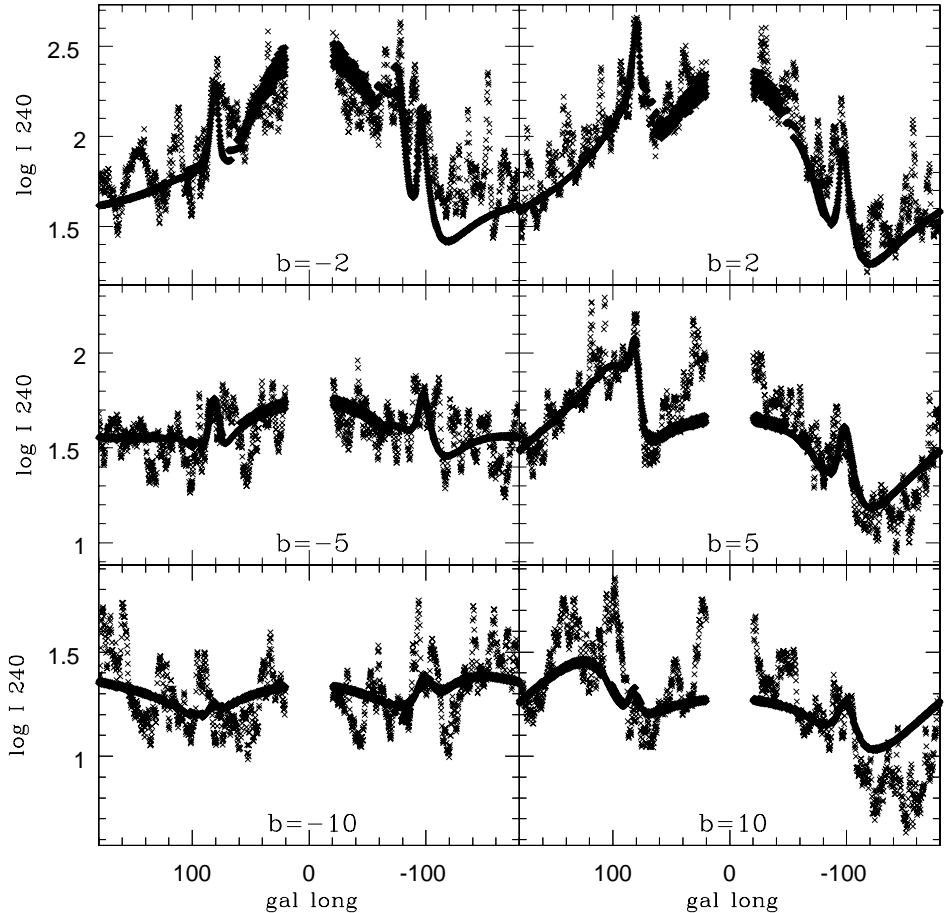}
\caption{
Observed and modeled $240 \micron$ emission profiles in 0.3\,deg wide
latitude bands at various galactic latitudes. 
}
\end{figure}

\section{Stellar distribution model}

In the NIR we assume negligible scattering, and that absorption is 
proportional to the dust column density. In this case we can write
\begin{equation}
\label{istar}
D_{\rm K}^{\rm mod}(l,b)
=\int_{\rm los} \eta(s) e^{-\tau(s)} ~ds + Q_{\rm K} ~,
\end{equation}
where $\eta$ is the flux density, $\tau$ the optical depth, and $Q_{\rm K}$
is again an isotropic term, though here not astrophysically justified.
The stellar flux density is comprised of an axisymmetric disk component 
with an exponential radial profile and a $\sech^2$ vertical profile. 
To this is added a spiral arm component. In the following section various
spiral arm geometries are applied, but in general the arms are given a double 
gaussian density profile in their cross section. 
A warp is also assigned to the stellar flux density,
but with a different amplitude than that in the dust.

The absorption is derived from the dust model. First the dust model
is rescaled, or corrected, to recover the observed FIR intensity along 
each line-of-sight. That is, along each line-of-sight a rescaling factor
is applied to {\em one} of the dust density components such that 
$\rho_j \rightarrow f_j \rho_j$, where the rescaling factor 
for the $j$th component is specified by
\begin{equation}
\label{rescale}
f_j = \frac{D_{240}^{\rm obs} - \sum_{i \neq j}D_i - Q_{240}}{D_j} ~.
\end{equation}
The component chosen is that which needs the least fractional change in
density to recover the observed FIR intensity, specifically, 
that which minimizes $|1 - f_j|$.
The resulting rescaled column density along each line-of-sight 
is then used to calculate the optical depth in Equation 3.

The parameters of the stellar distribution are adjusted to fit the model
K band emission to the observed, and 
at the same time the dust densities and emissivities are decoupled.
Figure 1 shows the resulting fit in the GP, and Figure 3 shows the emission
profiles at other galactic latitudes. At these wavelengths nonaxisymmetric 
structures are far from evident, and one might even conclude that the GP 
profile deviates from an axisymmetric profile due to absorption 
effects alone (Kent, Dame \& Fazio 1991).
However, the model shown in Figures 1 and 3 in fact has a two arm logarithmic
spiral. The next section will show fits using other (non)spiral models for 
comparison. 

Before going on it is worth pointing out one nonaxisymmetric feature that is
subtly but clearly present, namely the Galactic warp. The warp 
accounts for the skewness seen in the emission profiles at latitudes
$|b| \geq 5\deg$, as it starts within the Solar Circle (at $0.89 R_\odot$). 
The only way to reproduce such skewed profiles
is to introduce a local tilt to the stellar disk with respect to the
conventionally defined $b=0$ plane, which can also be done with
a large-scale tilt to the stellar disk (Hammersley et al. 1995). 

\begin{figure}[ht]
\plotone{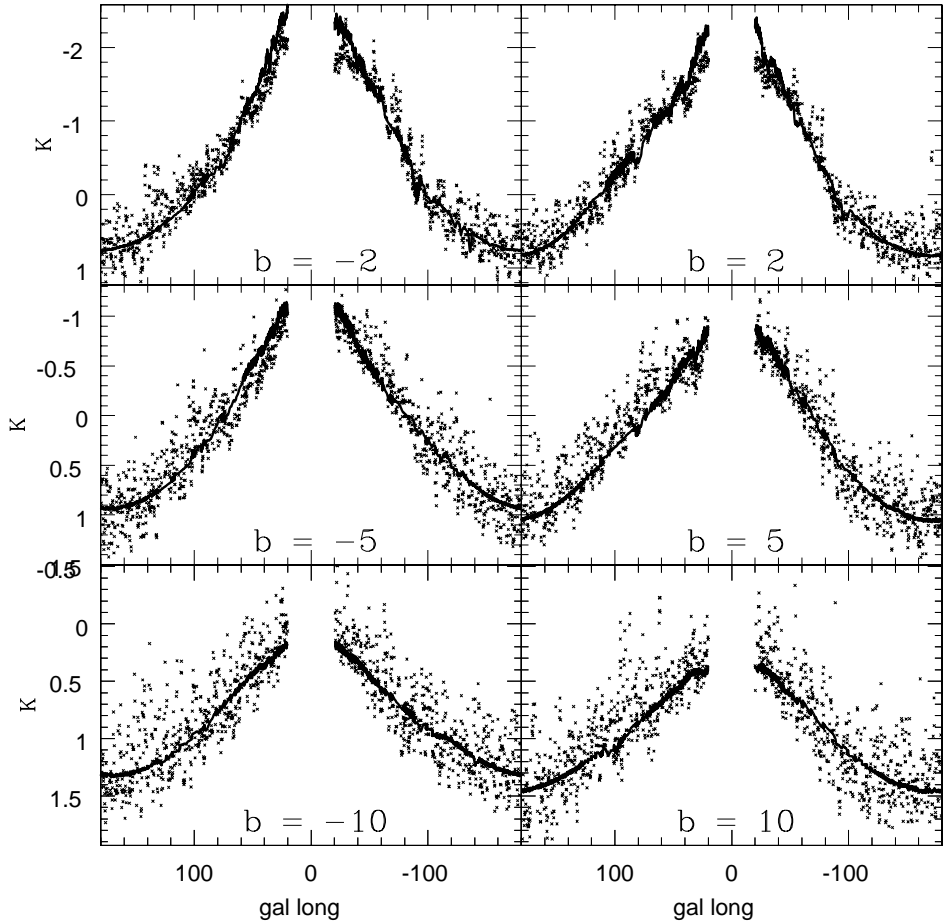}
\caption{
K band data and model profiles (solid lines) in $0.3\deg$ wide
latitude bands at various galactic latitudes.
}
\end{figure}

\section{Alternative spiral models}

\begin{figure}[ht]
\plotone{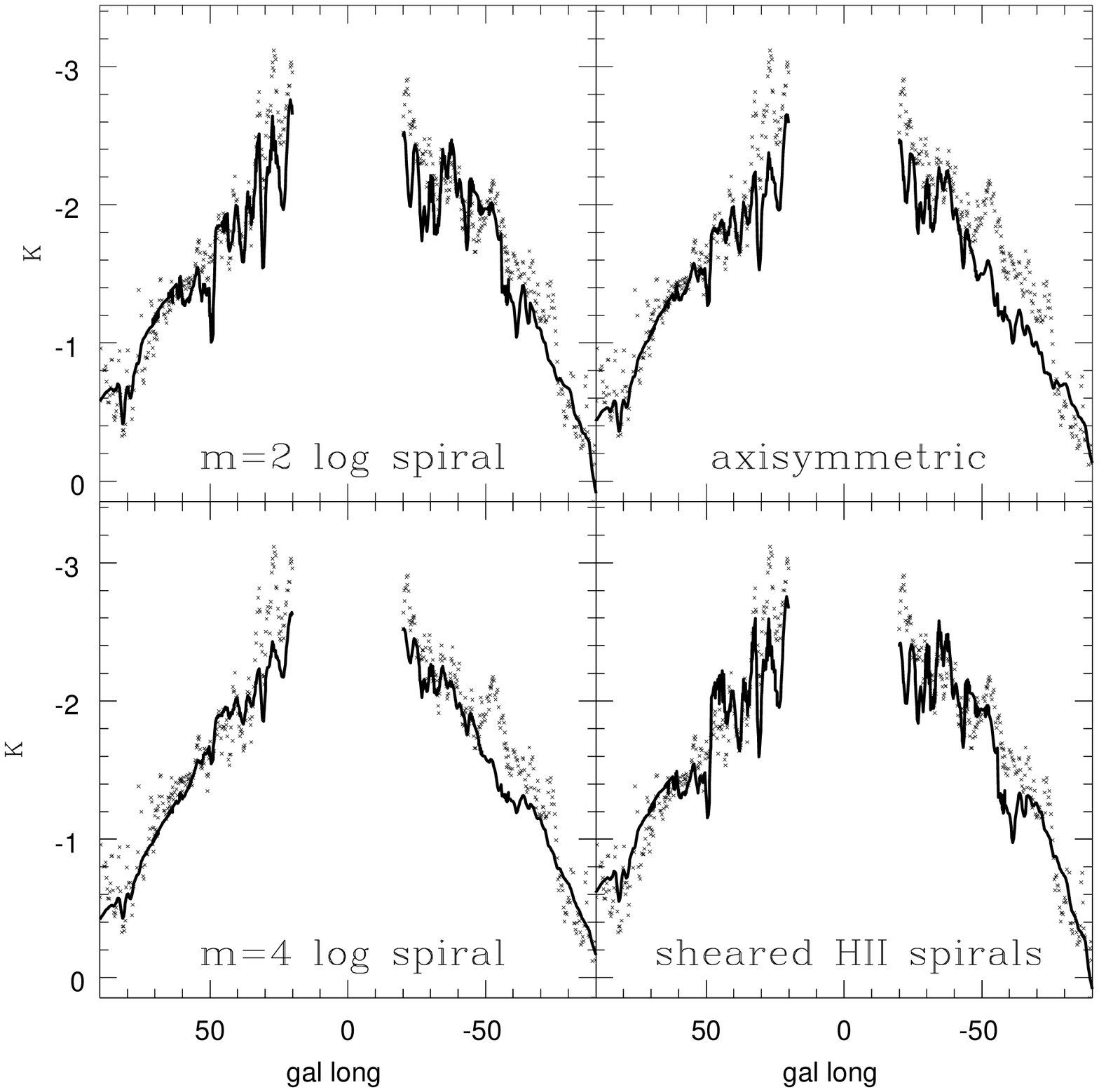}
\caption{
K band emission profiles within $0.17\deg$ of the GP 
for four spiral arm models against 
the observed emission profile, for galactic longitudes $|l|<90\deg$.
Models differ most in the direction of the Scutum arm tangent at
$\approx -55\deg$.}
\end{figure}

Two general spiral
geometries are here considered. The first is a logarithmic model whose pitch 
angle and phase are adjusted. Both $m=2$ and $m=4$ arm geometries were tried.
The second geometry is a sheared version of the HII arms adopted for the
dust. If the arms are assumed to have a fixed pattern in the presence of
a flat rotation curve, then the drift in azimuth is
\begin{equation}
\label{shear}
\phi_\tau = V_o \tau \left( \frac{1}{r} - \frac{1}{R_{\rm C}} \right),
\end{equation}
where $V_o$ is the circular rotation speed, $\tau$ the mean age of the
population, and $R_{\rm C}$ the corotation radius. In this formulation the
geometry is found by adjusting parameters equivalent to $V_o \tau$ and 
$R_{\rm C}$. A fourth model considered, for purposes of comparison, 
is a purely axisymmetric model. 

All these models give fits which differ only close to the plane of the Galaxy,
that is at $|b| < 3\deg$. The resulting disk scale lengths are found to 
be about $0.29 R_\odot$, except for the $m=4$ model which gives a
larger scale length of $0.37 R_\odot$, while the scale heights range from 
270 to 307\,pc and $Z_\odot$ from 12 to 15\,pc in the models (again assuming
$R_\odot = 8$\,kpc). Figure 4 shows the GP profiles for the 
four models just described. As can be immediately seen, the sheared arm model
and the $m=2$ logarithmic model give correspondingly good fits to the data,
while the $m=4$ and axisymmetric models fail to reproduce the observed profile
specifically in the directions of the tangents to the Scutum arm 
($l \simeq -55\deg$ and $30\deg$). No evidence of the Sag-Car arm is
seen at positive Galactic latitudes, as shown by the axisymmetric model.
The pitch angle for the $m=2$ log spiral is found to be $15.6\deg$, while 
that for the $m=4$ model is $12.4\deg$. For the sheared spiral model
a corotation radius of $R_{\rm C} = 0.83 R_\odot$ is arrived at, and a 
mean age of $\tau = 15$\,Myr is found if one adopts the values 200\,km/s 
and 8\,kpc for the circular speed and $R_\odot$ respectively.

\section{Conclusions}

For the K band emission two spiral models were found that reproduce the 
GP emission profiles, a $m=2$ logarithmic spiral model and a sheared 
version of the arms seen in the dust. This latter model is a 4 arm model,
however a reduction factor is applied to the Sag-Car arm that renders 
it significantly smaller than the other arms (a factor of 0.53 in two 
dimensions and in density). We thus conclude that in the
NIR a two arm spiral dominates the nonaxisymmetric emission.
This conclusion supports the earlier simple analysis of Drimmel (2000) 
based on the position of identified tangent features. However, the nature
of the arms are not clear; the open $m=2$ logarithmic arms suggest
an old population, as argued in Drimmel (2000), while the successful
sheared spiral model is consistent with the nonaxisymmetric emission 
arising from young K supergiants. Favoring this latter interpretation is 
the small scale height of the arms, which are not seen at $|b| > 3\deg$ 
from the GP. It is hoped that our future study including the J band will 
resolve this issue.

\acknowledgments

The COBE datasets were developed by the NASA
Goddard Space Flight Center under the guidance of the COBE Science
Working Group and were provided by the NSSDC.

\end{document}